# Statistics of diffusive and localized fields in the vortex core

Sheng Zhang and Azriel Z. Genack

Department of Physics, Queens College, The City University of New York, Flushing, NY 11367, USA

The statistics of the field structure in the vortex core surrounding phase singularities in random wave fields are measured and calculated for diffusive and localized waves. Excellent agreement is found between experiment and theory. The variation of phase with geometric angle is deterministic, depending only upon the eccentricity of elliptical intensity contours, $\varepsilon$, whose probability distribution is shown to be universal. The distribution of vorticity is shown to reflect both the vorticity distribution in the Gaussian limit and the mesoscopic distribution of total transmission.

The interference of multiply-scattered waves produces a quilt of bright spots associated with local intensity maxima. This speckled pattern is almost unavoidable when coherent sources are employed and its ubiquity is manifest whenever lasers beams are scattered. In addition to intensity maxima within the speckle pattern, there are isolated points of vanishing intensity at which the phase is undefined [1]. These are singular points in the phase from which equiphase lines radiate and across which the phase jumps by $\pi$ radians. Measurements [2] and computer simulations [3] have shown that speckle spots and phase singularities have the same density within random fields. Nye and Berry [1] showed that first order zeros of field at intensity nulls are generic elements of the speckle pattern, being stable under perturbation and omnipresent in random fields. Even though the fields are random, however, Berry and Dennis predicted a highly symmetric field structure near a phase singularity with elliptical intensity contours and circular flux current contours [4,5]. Due to the circular current flow at their core, phase singularities are also referred to as optical vortices. Within the vortex core, the field may be characterized by the eccentricity of the intensity contour, the vorticity of the current flow, and the angular variation of the phase. In recent measurements of the speckle pattern produced when a laser beam is transmitted though a sandblasted glass plate, Wang *et al.* [6] observed the predicted structure of intensity and current contours and inferred an exponential probability distribution of the eccentricities of the intensity contours. Though modifications of the statistics of the currents or flux within nonergodic speckle patterns in quantum and classical propagation due to long-range correlation, mesoscopic fluctuations, and wave localization have attracted considerable attention [7–16], the study of the structure of the speckle pattern has been largely confined to ergodic wave fields. The impact of mesoscopic correlation upon the evolution of the speckle pattern has been considered recently [17] but the unavoidable impact of correlation upon the statistics of field structures of static speckle patterns in the localization transition has not been considered.

In this Letter, we measure the statistics of field structures within the vortex core surrounding phase singularities for waves transmitted through mesoscopic samples. We show that the angular distribution of phase about a singularity can be fully described by the eccentricity of the elliptical intensity contours. The probability distribution of the eccentricity $\varepsilon$ and the vorticity, $\Omega$, $P(\varepsilon)$ and

$P(\Omega)$, within the vortex core are measured and calculated analytically for both diffusive and localized waves. We show that the statistics of eccentricity are universal; being the same for diffusive and localized waves and in accord with calculations by M. R. Dennis [5]. On the other hand, the statistics of vorticity directly reflect mesoscopic fluctuations with the variance of vorticity providing a direct measure of the degree of localization. The excellent correspondence between measurement and theory demonstrates that the statistics of the field deep within the vortex core can be measured by sampling the field using an antenna whose length is not negligible relative to the wavelength.

We measured the microwave field transmission coefficient through quasi-one dimensional random samples of alumina spheres contained in a copper tube for both diffusive and localized waves. The sample with diameter 0.95 cm and refractive index 3.14, which are embedded in Styrofoam shells of refractive index 1.04 to produce an alumina volume fraction of 0.068. [16] The copper tube has a diameter 7.0 cm and length of 61 cm with thin plastic end caps. The amplitude and phase of the field polarized along a wire antenna (4mm long and 0.5 mm wide) are measured with use of a vector network analyzer. The spatial distribution of the transmitted field over a range of frequencies is obtained by measuring the field spectra at each point of a 1-mm-square grid over the output plane. The spectra of speckle patterns are measured for diffusive waves from 14.7-15.7 GHz and for localized waves from 10-10.24 GHz in 1601 and 801 frequency steps, respectively. In each case, the frequency steps are approximately 1/7 of the field correlation frequency. New realizations of the random sample were created after the full speckle pattern was recorded by rotating and vibrating the sample tube. 40 and 58 configurations were measured for diffusive and localized waves, respectively. The wire antenna is chosen to be long enough to provides good signal noise ratio and a high degree of linear polarization state of the measured field. The measured field is therefore the sum of the field along the antenna rather than the polarized field at a single point. But since the measured signal is a random sum of partial waves, the sum of field along the antenna should still accurartely reflect the spatial structure and statistics of the random field. The accuracy of this cocnjecture can be judged by the extent of agreement between the measurements and theory presented below.

The field of a polarized speckle pattern for monochromatic radiation can be represented as,
$$\psi(x,y) = \xi(x,y) + i\eta(x,y) = A(x,y)\exp[i\varphi(x,y)],$$
where $\xi$ and $\eta$ are the real and imaginary parts of the field, respectively, $A$ is the amplitude and $\varphi$ is the phase. Phase singularities occur at the intersection of all equiphase lines, including contours with $\xi=0$ and $\eta=0$. Because the speckle patterns are produced by monochromatic radiation with magnitude of wave vector given by, $k = 2\pi v/c$, the bandwidth of the power spectrum in k-space $\Pi(k_x,k_y)$ is limited by the magnitude of the wavevector, i.e. $-k < k_x, k_y < k$. According to the two-dimensional sampling theorem [18], the lowest sampling resolutions required to perfectly reconstruct the full speckle patterns is $(1/2k) \times (1/2k)$, which corresponds to $1.5 \times 1.5$ mm$^2$ and $2.3 \times 2.3$ mm$^2$ for diffusive and localized waves, respectively. Our measurements on a 1-mm-square grid are therefore sufficiently dense to provide the speckle pattern with arbitrarily high resolution.

An example of a high resolution image of the field in the region surrounding a phase singularity is shown in Fig. 1a. Near the origin, close to the phase singularity, the contours of intensity, $I = A^2$, are ellipses while the contours of current, $J = |\vec{J}| = |I\nabla\varphi|$, are circles as predicted [4] and has been seen in recent optical measurements [6].

The vorticity associated with current can be defined as, $\vec{\omega} \equiv \tfrac{1}{2}\nabla \times \vec{J}$, so that the magnitude of vorticity in the *x-y*-plane is $\omega = |\vec{\omega}| = |\xi_x \eta_y - \eta_x \xi_y|$. Here the subscripts denote the direction along which the derivatives of field components are taken, e.g. $\xi_x \equiv \partial\xi/\partial x$. In the region close to a phase singularity, the magnitude of the current increases linearly with the distance from the singularity [4,6], $J(r) \approx \Omega r$, where, $\Omega = \omega(0)$. The equal spacing of contours with equal current increments seen in Fig. 1(a) indicates that the current is proportional to the radius. The vorticity at the singularity may be expressed as, $\Omega = \lim_{r \to 0} J(r)/r$.

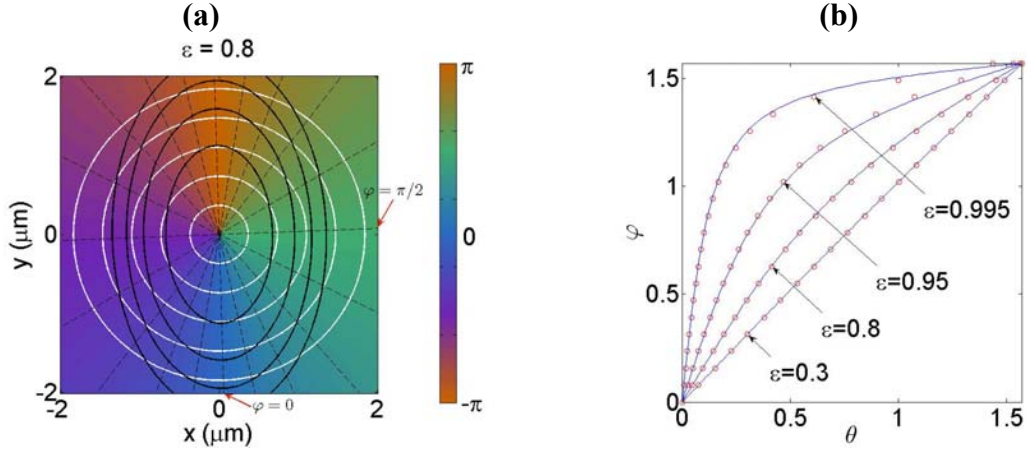

**Fig. 1. (a) Core structure of a phase singularity. Phases from −π to π are represented by a continuous color map. The dashed lines are equiphase lines with phase difference of π/10. Circles (white) are current contours, while ellipses (black) are intensity contours. (b) Relation between $\varphi$ and $\theta$ for given values of eccentricity. Since the phase contours are symmetric with respect to reflection about the minor and major axes of the ellipse, only the region $\varphi, \theta \in \left[0, \dfrac{\pi}{2}\right)$ is shown.**

The intensity contours close to a phase singularity are ellipses since the intensity at a small distance r from the singularity can be expressed by the Taylor expansion,

$$I(\vec{r}) \approx |\vec{r} \cdot \nabla \psi(0)|^2 = [\vec{r} \cdot \nabla \xi(0)]^2 + [\vec{r} \cdot \nabla \eta(0)]^2. \qquad (1)$$

The major and minor radii of the ellipse with *I*=1 are, respectively,

$$a = \frac{1}{\sqrt{2}\omega(0)}\sqrt{G(0) + \sqrt{G(0)^2 - 4\omega(0)^2}},$$
$$b = \frac{1}{\sqrt{2}\omega(0)}\sqrt{G(0) - \sqrt{G(0)^2 - 4\omega(0)^2}}. \qquad (2)$$

where, $G \equiv |\nabla \psi|^2 = (\nabla \xi)^2 + (\nabla \eta)^2$. The eccentricity $\varepsilon = \sqrt{1 - b^2/a^2}$ of this ellipse can therefore be expressed as [4],

$$\varepsilon = \frac{1}{\sqrt{2}\omega(0)}\left[G(0)^2 - 4\omega(0)^2\right]^{1/4}\sqrt{G(0)-\sqrt{G(0)^2-4\omega(0)^2}}\,. \qquad (3)$$

The phase change in a circuit around any of the singularity in our experiments is $\pm 2\pi$. We see in Fig. 1(a) that all phase contours radiate from the singularity as straight lines. Choosing the phase of the equiphase line along one of the semi-major axes of the elliptical intensity contour to be 0, the geometric angle $\theta$ between the 0 and the $\varphi$ equiphase lines is taken in the counterclockwise sense. Equation (1) can then be written as,

$$I(\vec{r}) = r^2\left(\frac{\cos^2\theta}{a^2} + \frac{\sin^2\theta}{b^2}\right), \qquad (4)$$

where $\vec{r} = r\exp(i\theta)$. Since $J(\vec{r}) = I(\vec{r})|\nabla\varphi(\vec{r})|$ and $|\nabla\varphi(\vec{r})| = \frac{1}{r}\frac{d\varphi(\theta)}{d\theta}$, we find using Eq. (4) and the relation, $J(r) \approx \Omega r$,

$$\frac{d\varphi(\theta)}{d\theta} \approx \frac{\Omega}{\cos^2\theta/a^2 + \sin^2\theta/b^2}. \qquad (5)$$

From Eq. (2) we find, $ab\Omega = 1$. Integrating Eq. (5), $\varphi(\theta) = \int_0^\theta \frac{d\varphi(\theta')}{d\theta'}d\theta'$, gives a relation between the relative phase angle $\varphi(\theta)$ and the geometrical angle $\theta$, which depends only upon $\varepsilon$,

$$\tan[\varphi(\theta)] = \tan(\theta)/\sqrt{1-\varepsilon^2}\,. \qquad (6)$$

Measurement of $\varphi(\theta)$ around singularities with different values of $\varepsilon$ are seen in Fig. 1(b) to be in excellent agreement with Eq. (6).

The probability distribution of $\varepsilon$ near singularities was inferred by Wang, *et al.* [6] from measurements of speckle patterns of laser beams passed through a roughened glass plate. The distribution was approximately an exponential function in the region, $0.4 < \varepsilon < 1$, over which statistics were available.

Our measurements were carried in mesoscopic samples in which fluctuations of transmission are greatly enhanced. This can be seen from the probability distribution of total transmission normalized by its ensemble average, $s_a = T_a/\langle T_a\rangle$, shown in Fig. 2. Here $T_a$ is obtained by summing the intensity over the output surface. The distribution is significantly broadened for localized waves [13,15]. The extent of photon localization can be characterized by var($s_a$) with the localization threshold at var($s_a$) = 2/3 [15,16]. In the diffusive limit, $P(s_a)$ is a delta function at $s_a = 1$. In the two frequency ranges studied here, var($s_a$) = 0.14 and 3.0. The probability distributions $P(s_a)$ measured for diffusive and localized waves are shown in Fig. 2 and agree the expressions given in [12] as interpreted in [13,15].

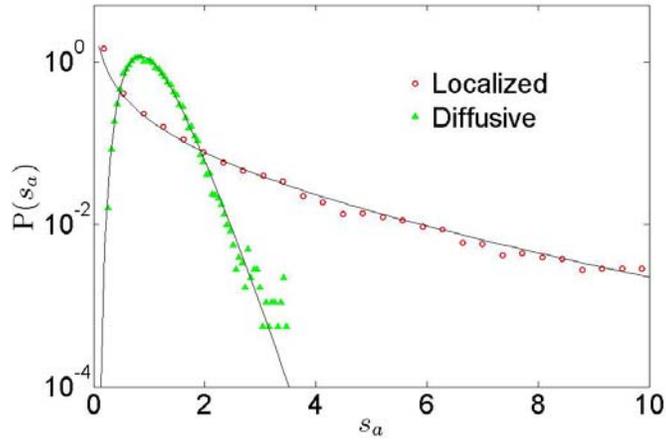

**Fig. 2. Probability distributions of normalized total transmission $s_a$. Green triangles and red circles are experimental data for diffusive and localized waves, respectively. Solid lines are theoretical calculations.**

Our measurements of the probability distributions of eccentricity $P(\varepsilon)$ for over $10^5$ singularities for diffusive and localized waves are presented in Fig. 3. We find that the distributions are identical. They are close to exponential for $\varepsilon > 0.4$, but fall much more sharply for $\varepsilon < 0.4$.

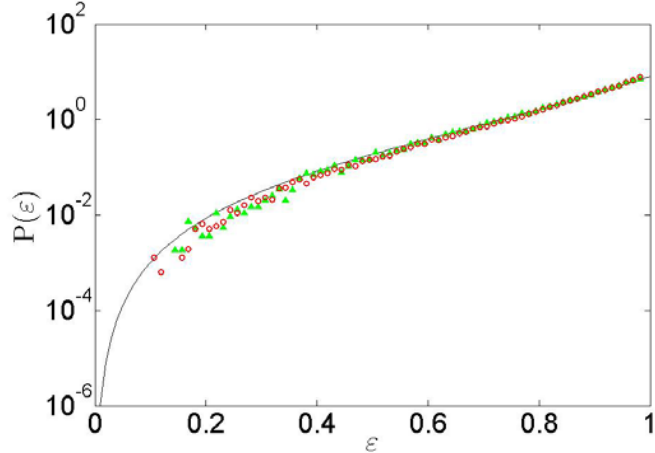

**Fig. 3. Probability distributions of $\varepsilon$. The solid line is Eq.(7) calculated for Gaussian fields.**

Measurements of $P(\varepsilon)$ are compared to the expression derived for Gaussian random fields by Dennis [5],

$$P(\varepsilon) = \frac{8\varepsilon^3}{(2-\varepsilon^2)^3} . \tag{7}$$

Equation (7) is plotted as the solid curve in Fig. 3 and seen to be in good agreement with measurements.

We next compare the probability distributions of vorticity at the singularity for diffusive and localized waves. In order to compare these distributions, the vorticity is normalized by its ensemble average value, $\tilde{\Omega} = \Omega/\langle\Omega\rangle$, and plotted in Fig. 4. $P(\tilde{\Omega})$ is seen to be significantly wider for localized waves.

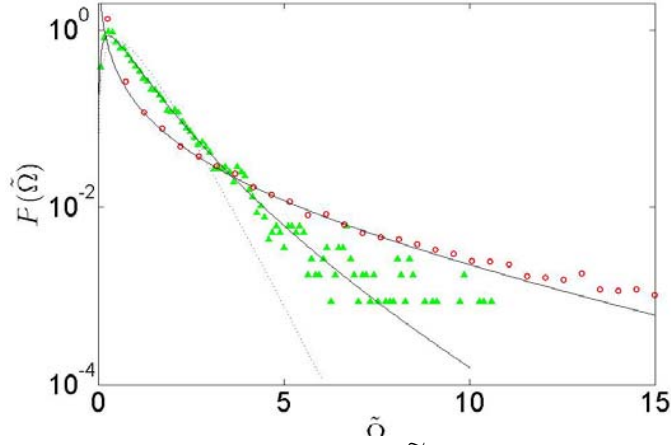

**Fig. 4. Probability distributions of vorticity $\tilde{\Omega}$. Solid lines are Eq.(13) and dashed line is $P(\Omega_G/\langle\Omega_G\rangle)$ in Eq.(12).**

These measurements will be compared to the theoretical calculation of $P(\Omega_G/\langle\Omega_G\rangle)$ for Gaussian random fields. The subscript $G$ indicates Gaussian statistics. Berry and Dennis pointed out that since the vorticity $\omega$ is the Jacobian of the transformation from $\xi$ and $\eta$ to Cartesian coordinates, the density of singularities is given by, $d = \langle\delta(\xi)\delta(\eta)\omega\rangle = \frac{K_2}{4\pi}$, where $K_2$ is the second moment of the distribution of the magnitude of the transverse component of the $k$-vector [4]. The average of any quantity $f$ associated with singularities can be given by, $\langle f \rangle_s = \frac{1}{d}\langle\delta(\xi)\delta(\eta)\omega f\rangle$, where $\langle\ \rangle_s$ denotes the average over all singularities [4]. Therefore $P(\Omega_G)$ can be expressed as,

$$P(\Omega_G) = \frac{1}{d}\langle\delta(\xi)\delta(\eta)\omega\delta(\omega-\Omega_G)\rangle. \qquad (8)$$

Since the field, $\psi = \xi + i\eta$, is a complex Gaussian random variable, $\langle\delta(\xi)\rangle = \langle\delta(\eta)\rangle = 1/\sqrt{2\pi}$. Furthermore, since $\omega$ is independent of the gauge of phase set by the in- and out-of-phase components of the fields $\xi$ and $\eta$, Eq. (8) can be expressed as,

$$P(\Omega_G) = \frac{1}{2\pi d}\int_0^\infty d\omega\,\omega\,\delta(\omega-\Omega_G)P(\omega). \qquad (9)$$

Using the joint probability distribution $P(\omega,G) = \frac{2}{K_2^2}\exp\left(-\frac{G}{K_2}\right)\Theta(G-2\omega)$ given in [4], where $\Theta$ denotes the unit step function, $P(\omega)$ may be expressed as,

$$P(\omega) = \int_0^\infty P(\omega,G)dG = \frac{2}{K_2}\exp\left(-\frac{2\omega}{K_2}\right), \qquad (10)$$

Substituting Eq.(10) into Eq.(9) gives,

$$P(\Omega_G) = \frac{4\Omega_G}{K_2^2}\exp\left(-\frac{2\Omega_G}{K_2}\right) \qquad (11)$$

Using this result, we find that $\langle\Omega_G\rangle = K_2$ and,

$$P(\tilde{\Omega}_G) = 4\tilde{\Omega}_G \exp(-2\tilde{\Omega}_G), \qquad (12)$$

for the probability distribution of normalized vorticity, $\tilde{\Omega}_G = \Omega_G/\langle\Omega_G\rangle$. This result for the Gaussian limit is shown as the dashed curve in Fig. 4.

To examine whether the Gaussian and mesoscopic effects can be separated, we consider the probability distribution of the vorticity normalized by the total transmission in each sample configuration, $\Omega' = \Omega/s_a$. The resulting probability distribution, $P(\tilde{\Omega}' = \Omega'/\langle\Omega'\rangle)$, shown in Fig. 5, is the same for diffusive and localized waves and in good agreement with $P(\tilde{\Omega}_G)$ given in Eq.(12). This implies that the normalized vorticities, $\Omega' = \Omega/s_a$, and, $\tilde{\Omega}' = \tilde{\Omega}/s_a$, are statistically independent of fluctuation in $s_a$. Thus the $P(\tilde{\Omega})$ may be expressed as a mixture of $P(\tilde{\Omega}_G)$ given in Eq. (12) and $P(s_a)$,

$$P(\tilde{\Omega}) = \int_0^\infty \frac{ds_a}{s_a}\frac{4\tilde{\Omega}}{s_a}\exp(-2\tilde{\Omega}/s_a)P(s_a). \qquad (13)$$

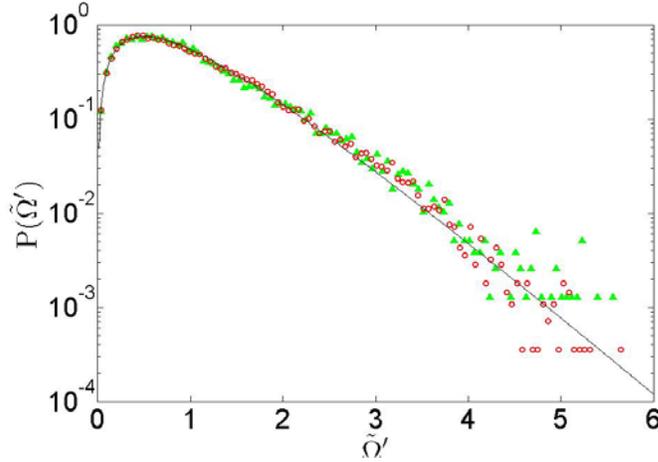

**Fig. 5. Probability distributions of $\tilde{\Omega}'$. The solid line is Eq. (12).**

Calculation of $P(\tilde{\Omega})$ based on measurements of $\text{var}(s_a)$ for diffusive and localized waves and the theoretical expression of $P(s_a)$ in Ref. [12, 13] are shown as the solid curves in Fig. 4 and seen to be in excellent agreement with experimental results.

From Eq. (12), we find,

$$\text{var}(\tilde{\Omega}) = \frac{3}{2}\text{var}(s_a) + \text{var}(\tilde{\Omega}') = \frac{3}{2}\text{var}(s_a) + \frac{1}{2}, \qquad (14)$$

where $\text{var}(\tilde{\Omega}') = 1/2$ is the result for a Gaussian random field. Thus, $\text{var}(\tilde{\Omega})$ represents the sum of ergodic and nonergodic fluctuations of the speckle pattern.

In conclusion, the excellent agreement between experiment and theory demonstrates that it is possible to measure wave statistics deep within the vortex core. The statistics of vorticity provide a measure of mesoscopic fluctuations and photon localization. In contrast, the statistics of $\varepsilon$ represent the geometry of the speckle pattern and are universal.

We thank Bing Hu and Patrick Sebbah for experimental help at an early stage of this work. This research was supported by the NSF under grant number DMR-0538350.